\documentclass[conference]{IEEEtran}

\usepackage{amssymb, amsmath}
\usepackage{macros}
\usepackage{wrapfig}
\usepackage{framed}
\usepackage{paralist}
\usepackage{pifont}
\usepackage{hyperref,breakurl}

\newtheorem{thm}{Theorem}
\newtheorem{theorem}[thm]{Theorem}

\newtheorem{defin}{Theorem}
\newtheorem{definition}[defin]{Definition}
\newtheorem{prob}{Theorem}
\newtheorem{problem}[prob]{Problem}

\newtheorem{sys}{Theorem}
\newtheorem{system}[sys]{System}

\newcommand\tick{\ding{51}}
\newcommand\crossMark{\ding{54}}

\begin{document}
\title{Counterexample Guided Synthesis of Switched Controllers for Reach-While-Stay Properties}
\author{Hadi~Ravanbakhsh

and~Sriram~Sankaranarayanan
}

\maketitle

\begin{abstract}

We introduce a counter-example guided inductive synthesis (CEGIS)
framework for synthesizing continuous-time switching controllers that
guarantee reach while stay (\RWS) properties of the closed loop
system. The solution is based on synthesizing control Lyapunov
functions (CLFs) for switched systems, that yield switching
controllers with a guaranteed minimum dwell time in each mode. Next,
we use a CEGIS-based approach to iteratively solve the resulting
quantified exists-forall constraints, and find a CLF. We introduce
refinements to the CEGIS procedure to guarantee termination, as well
as heuristics to increase convergence speed. Finally, we evaluate our
approach on a set of benchmarks ranging from two to six state
variables, providing a preliminary comparison with related tools.  Our
approach shows significant speedups, thus demonstrating the promise of
nonlinear SMT solvers for synthesizing provably correct switching
control laws.
\end{abstract}



\IEEEpeerreviewmaketitle

\section{Introduction}

\IEEEPARstart{I}{n} this paper, we study the problem of automatically synthesizing
continuous-time switching controllers for
ensuring \emph{Reach-While-Stay} (\RWS) properties of polynomial
systems. \RWS properties specify a set of \emph{goal states} $G$ and a
set of safe states $S$. The controller synthesis consists of
synthesizing a winning region $ W \subseteq S$ and $W$
such that the system initialized inside $W$ is guaranteed to stay
inside $W$ \emph{until} a goal state in $G$ is eventually
reached. Additionally, to ensure that the winning region is large
enough, we specify an initial set $I$ and require that $I \subseteq
W$. Such properties commonly arise in many situations such as
stabilizing the output of a system to a goal region, while ensuring
that the intermediate ``transients'' stay within safe bounds, or
enable an autonomous vehicle to reach a target while staying away from
obstacles. Our approach considers switched system plant models with
finitely many control modes and continuous state variables whose
dynamics in each mode are described by ODEs. The goal of the
controller synthesis is to find a switching controller that chooses a
control mode, given the current mode and continuous state.

Our approach synthesizes a control Lyapunov function (CLF), which can be made to
decrease along the traces of the closed loop system, guaranteeing that the
traces reach a designated, desirable region while staying in the safe region. As
such, finding a CLF yields a switching function that simply chooses an
appropriate control mode that ensures its decrease. However, for continuous-time
switching, we are faced with the problem of \emph{zenoness} caused by the
controller switching infinitely often in a finite time interval, and thus,
preventing time from diverging. Therefore, we provide sufficient conditions on
the switching strategy that ensure that the resulting switching function
respects a minimum dwell time for each control mode.

The synthesis procedure iteratively searches for a CLF using
Satisfiability Modulo Theory (SMT) solvers through a
well-known procedure for program synthesis called
\emph{counter-example guided inductive synthesis}
(CEGIS)~\cite{SolarLezema/08/Program,solar2006combinatorial,alur2013syntax}. Whereas
CEGIS was originally proposed for synthesizing unknown parameters for
programs (called \emph{sketches}) so that assertions (safety
properties) in the program are satisfied by all executions, we propose
to reuse the basic insights for synthesizing
controllers. 
Since the search space is infinite (and continuous), there is no
guarantee that the process terminates. We show an adaptation of
CEGIS to our setting that ensures eventual termination of the CEGIS
algorithm.

We provide an implementation of the CEGIS approach for synthesizing
controllers using the SMT solvers Z3 for linear
arithmetic~\cite{DeMoura+Bjorner/08/Z3} and the dReal
$\delta-$satisfiability solver for nonlinear arithmetic
constraints~\cite{DBLP:conf/cade/GaoKC13}. The evaluation shows the
ability of our approach to effectively synthesize switching
controllers with guaranteed minimal dwell time for $20$ benchmark
systems drawn from the related literature. On the other hand, the high
complexity of nonlinear arithmetic decision procedures cause our
approach to fail on 2 out of the 20 benchmarks
attempted. Nevertheless, we provide a preliminary comparison that
suggests that our approach is quite competitive with other
state-of-the-art approach for the synthesis of controllers to
guarantee temporal logic objectives.  However, we are exploring
relaxations for the non-linear constraints using well-known schemes
such as SOS programming~\cite{Parillo/2003/Semidefinite}. An important
technical limitation of these relaxations that prevents their direct
use in this paper lies in the lack of useful witnesses in case a given
constraint is satisfiable.  The contributions of this paper are
summarized below: (A) We synthesize minimum dwell-time enforcing
controllers that guarantee \RWS from CLFs. (B) We adapt the well-known
CEGIS algorithm to discover CLFs for polynomial switched
systems~\cite{SolarLezema/08/Program,solar2006combinatorial,alur2013syntax}.
(C) We show how the CEGIS search for candidate CLFs can be modified to
guarantee finite termination.  (D) We employ a heuristic to find
better witness points that significantly improve the proposed
approach.  (E) We provide an experimental evaluation on a number of
interesting benchmarks that demonstrates the promise of our approach,
as well as its limitations.

\subsection{Related Work}
\paragraph{Verification:} The stability of 
hybrid systems has been studied widely. Lyapunov functions
remain a simple, yet powerful, approach for proving various forms of
stability. The problem of synthesizing Lyapunov functions has been
approached using ideas such as SOS programming that reduces the
conditions for a Lyapunov function for a given system to a
semi-definite optimization problem
~\cite{papachristodoulou2002construction,Parillo/2003/Semidefinite,Tibken/2000/Estimation}.
Lyapunov function approach naturally extends to liveness properties
(such as \RWS).

\paragraph{Synthesis:} Beyond verification, much work has focused on designing
correct-by-construction controllers for various liveness properties,
especially stability.  The problem for synthesis is generally much
harder than verification. A common approach to synthesizes
\emph{control Lyapunov functions} (CLFs) whose values can be decreased
at each time instant through an appropriate control
input~\cite{artstein1983stabilization}. For continuous-time plants,
CLFs can yield an associated static feedback law that guarantees
stability under some necessary conditions originally proposed by
Artstein~\cite{artstein1983stabilization}.  The problem formulation
for synthesizing CLFs yields NP-hard, bilinear matrix inequalities (BMI).
The  BMI problem is  solved directly using gradient descent
~\cite{henrion2005solving} or using a heuristic such as V-K iteration
~\cite{el1994synthesis} (or elsewhere called policy iteration
~\cite{gaubert2007static}), which is often susceptible to failures due
to local minima.  Tan et. al.~\cite{tan2004searching} formulate these
conditions as a BMI and use off-the-shelf approaches to tackle the
resulting BMIs.  Rifford~\cite{rifford2000existence} discusses the
converse results on control Lyapunov functions, i.e. if a system is
globally asymptotically controllable, then there exists a locally
Lipschitz control Lyapunov function. This justifies the use of
Lyapunov function based methods. 

\paragraph{Switched Controllers:} 
In this article, the problem is to find a switching logic such that
the closed loop system satisfies a \RWS \ property. A large
volume of work on switched system has focused on linear
switched systems and the use of linear matrix inequalities to find
controllers. Details are available from the textbook by
Liberzon~\cite{liberzon2003switching}, and the survey articles by Lin
and Ansaklis~\cite{lin2009stability,lin2014hybrid}. Our approach here
considers the continuous-time switched systems of more general
polynomial dynamical systems. Furthermore, our focus is on synthesis
to guarantee a minimum dwell time in each switching mode, and the use
of CEGIS to find CLFs. These aspects are, to the best of our
knowledge, unique to this work.

Another approach to controller synthesis is proposed by Taly et
al.~\cite{taly2010switching}. They consider the problem of
synthesizing switching conditions for hybrid systems, so that the
resulting system guarantees safety and liveness properties. The
proposed method proves reachability by finding some progress
certificates similar to Lyapunov functions. They reduce their
synthesis to solving a system of nonlinear constraints. Our work here
differs in the certificates used to prove desired property (these
certificates are much simpler in our method) and the process of
finding such certificate.

Dimitrova and Majumdhar investigate proof systems for solving general
parity games on continuous state-spaces using Lyapunov-like
functions~\cite{Dimitrova+Majumdar/2014/Deductive}. Their approach
subsumes \RWS properties. However, they do not provide an approach to
synthesize these functions.  The CEGIS approach presented here
is a good candidate for such a mechanization, and the combination
will be investigated as part of future work.


 Another paradigm for synthesizing controllers is to define an abstract
 system and find a simulation (or approximate bisimulation) relation
 between the abstract system and the original system.  These
 approaches are able to handle more general specifications (usually a
 sub-class of LTL) and they are not restricted to liveness properties, per se.
 The PESSOA tool ~\cite{Mazo+Others/2010/PESSOA} uses finite
 abstraction to discretize a continuous-time system, and solves a
 completely discrete problem. One problem with this approach is the
 number of the abstract states, which can grow very large if we want
 them to be precise. Recently, C{\'a}mara
 et. al.~\cite{camara2011synthesis} proposed multi-scale abstraction
 to keep the number of states small and subsequently, Nilsson
 et. al.~\cite{nilssonincremental} proposed a CEGAR-based approach to
 refine the abstraction, whenever it is needed to avoid large number
 of abstract states. Our approach does not directly partition the
 state-space.  On the other hand, the nonlinear SMT solvers such as
 dReal that are used in the CEGIS approach implicitly partition the
 state-space during the search for a CLF.  However, such a
 partitioning is adaptive and is guided by the formula whose
 satisfiability is being decided. The preliminary evaluation provided
 shows that our approach can potentially be much faster in terms of
 time.

\paragraph{Fast Switches}
Most of aforementioned works consider discrete-time feedback for switched
systems. When the feedback is continuous, extra care should be taken
for infinitely fast switches. This phenomenon is common in many types
of control, including sliding mode
control~\cite{lin2007switching,geromel2006stability}.  Asarin
et. al.~\cite{asarin2000effective} propose another method for
enforcing min-dwell time property for finite-abstraction based
synthesis.  Taly et. al. ~\cite{taly2010switching} use ``Progress
Invariants'' to prove min-dwell time properties (for each switch the
value of Lyapunov function should decrease at least $\epsilon > 0$
unit).  Here, we develop a simpler strategy
to guarantee a minimum dwell time.

\paragraph{CEGIS:} 
The CEGIS framework has been used widely for solving $\exists \forall$ formulae
in synthesis problems ~\cite{alur2013syntax}. The idea here is to find a
candidate solution based on some finite number of examples.
Although CEGIS was first proposed in the computer science
literature by Solar-Lezama et al.~\cite{SolarLezema/08/Program}, variants
of this approach are not unknown to the hybrid systems community. 
This strategy has been mainly used to find a solution for $\exists \forall$ formulae
to solve parameter synthesis problems in programming languages
(~\cite{SolarLezema/08/Program,alur2013syntax,solar2006combinatorial} to mention few)
and hybrid systems
~\cite{frehse2008counterexample,yordanov2008parameter}.
Topcu et al. consider a simulation-based approach for finding maximal
region of attraction for continuous
systems~\cite{Topcu+Packard+Seiler+Wheeler/2007/Stability}. They employ
a CEGIS-like approach that avoids solving a BMI through sampling
finitely many witness points that are likely to belong to the region
of attraction.  A LMI is used to search for a Lyapunov function that
includes these witness points. Also Kapinski et. al.~\cite{kapinski2014simulation}
employ a CEGIS approach for synthesizing Lyapunov functions 
based on simulation results (initiated from witness points). 
In contrast, our approach considers
switched systems and focuses on synthesizing CLFs. We also do not
perform any sort of simulations for the witness points in our
approach.

\section{Preliminaries}
Let $\mathbb{N}$, $\reals$ and $\reals^+$ denote the set of natural
, real and nonnegative real numbers respectively.  Let $\vzero$ be
the zero vector of proper size.  For $n \in \mathbb{N}$ and
real number $\delta > 0$, let $\mathcal{B}_{\delta}(\vx_c)$ be a ball
with radius $\delta$ and $\vx_c$ as its center
($\mathcal{B}_{\delta}(\vx_c) = \{\vx \ | \ \norm{\vx - \vx_c} \leq
\delta\}$). For a set $S$, let $\partial S$ and $int(S)$ be 
boundary and interior of $S$, respectively.
  Let $\Poly[\vx]$ denote the set of all polynomials
involving variables in $\vx$, wherein each polynomial is written as a
finite sum $p: \sum_{\alpha \in \mathbb{N}^n} c_{\alpha}
\vx^{\alpha}$, where the multi-index $\alpha$ is used to denote a
monomial $\vx^{\alpha}$ and $c_{\alpha} \in \reals $ is a
coefficient. A \emph{template polynomial} over coefficients $C$ is a
polynomial $F(\vx, \vc): \sum_{\alpha \in \mathbb{N}^n} c_{\alpha}
\vx^{\alpha}$ whose coefficients are parameterized by a set of
template variables $c_{\alpha} \in C$.  Given a function $f: \reals
\rightarrow \reals$, $f^-(t)$ denotes the left limit: $\lim_{s
  \rightarrow t} f(s)$ and $f^+(t)$ denotes the right limit: $\lim_{t
  \leftarrow s}f(s)$.  As a convention, let $\dot{f}(t)$ denote the
\emph{right derivative} of the function: $\lim_{h \rightarrow 0}
\frac{ f(t + h) - f(t)}{h}$ at $x=t$. If $f$ is differentiable then
$\dot{f}$ coincides with its derivative.

\begin{wrapfigure}{r}{0.2\textwidth}
\begin{center}
\includegraphics[width=0.2\textwidth]{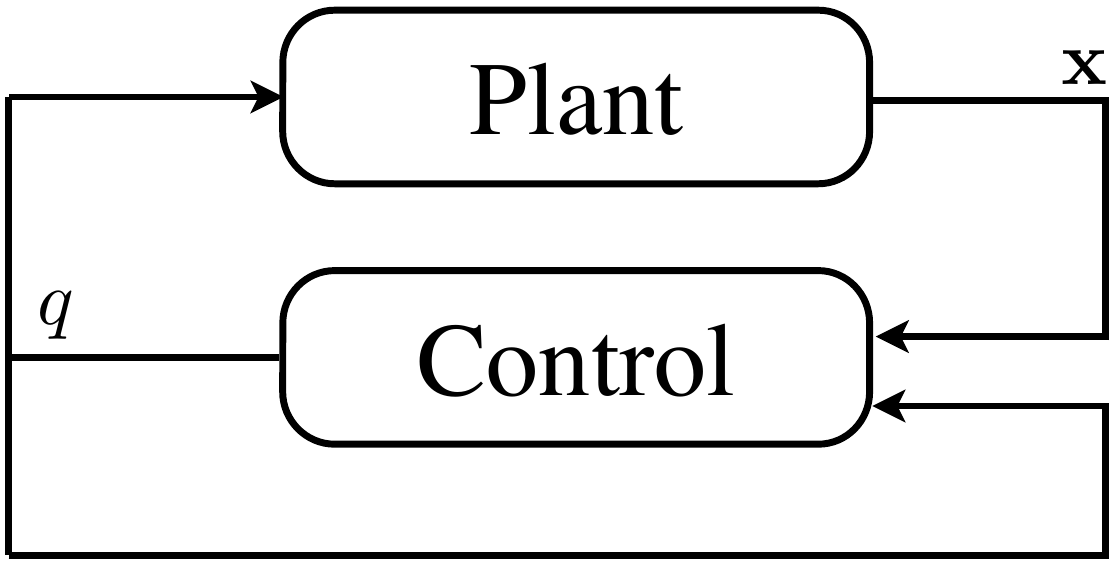}
\end{center}
\vspace*{-0.3cm}
\caption{The closed loop model of the plant and the controller.}\label{fig:feedback}
\end{wrapfigure}

\paragraph{System Model:} We first discuss the system model for our
controller synthesis problem. The system has a \emph{plant model}
which describes the physical environment with continuous
dynamics. Also, the system has a controller which provides a mode for
the plant (Figure ~\ref{fig:feedback}). 

The plant has continuous state variables $\vx \in \reals^n$.  The
controller continuously chooses a mode from a finite set of possible
\emph{control modes} $q \in Q$, wherein the dynamics of the continuous
state variables may depend on the chosen control mode $q$.
We now define our plant models.

\begin{definition}[Switched Polynomial System]
\label{def:plant}
A \emph{switched polynomial system} is a tuple $\Psi : (Q, X, f)$,
consisting of (A) \emph{Continuous state-space}: $X \subseteq
\mathbb{R}^n$ ($n$ is the number of continuous state variables); (B) A
finite set $Q$ of (control) modes; and (C) A map from each mode $q \in
Q$ to a polynomial vector field $f_q \in \Poly[\vx]^n$, specifying its
dynamics.
\end{definition}

The controller is modeled as a memoryless state feedback switched controller.
\begin{definition}[Switching Controller]
  Given a plant $\Psi: (Q,X,f)$, a switching controller $\ctrl$ is
  specified by a function  $\ctrl(q,\vx)$ that maps each 
  current mode $q \in Q$ and plant state $\vx \in
  X$ to a next mode $\hat{q} \in  Q$.
\end{definition}

A control implementation conforms to this specification by choosing
the next mode specified in $\ctrl(q,\vx)$.  
\begin{definition}[Closed Loop System]
  The composition of the plant $\Psi$ and a controller $\ctrl$ yields
  a \emph{deterministic switched system} $\Phi(X, Q, f, G)$ with continuous
  variable $\vx$, modes $Q$, 
  dynamics given by $f_q$ in each mode $q \in Q$.  The transition
  from mode $q$ to $\hat{q}$ has the guard set 
  $G_{q,\hat{q}}:\ \{\vx | \ctrl(q,\vx) = \hat{q}\}$

\end{definition}

The set of traces of the closed-loop switched system represents
all executions of the switched system that respects
the plant dynamics and switches according to the controller specification. 
Formally,  a  \emph{trace} $\tr: \reals^+ \rightarrow Q \times X$  is a function
 mapping  time
$t \in \reals^+$ to the mode and state of the plant at that time.
Let $\tr_X$ ($\tr_Q$) denote  the projections of the trace $\tr$ onto the sets
$X$ (and $Q$). For a trace to be valid, it  must satisfy the following conditions:
\begin{itemize}

\item The switching times
$ \mathsf{SwitchTimes}(\tr):\ \{ t \in \reals^+\ |\ \tr_Q^+(t) \not= \tr_Q^-(t) \} $
form a finite, or countably infinite set.

\item For all \emph{non-switching times} $t \in \reals^+ \setminus   \mathsf{SwitchTimes}(\tr)$,
writing $q:\ \tr_Q(t)$, we have
$\tr_X$ is differentiable at $t$ and $\frac{d \tr_X}{dt} = f_q(\tr_X(t))$.

\item For all \emph{switching times} $t \in \reals^+ \cap \mathsf{SwitchTimes}(\tr)$, writing
$q: \tr_Q^-(t)$ and $\hat{q}:\ \tr_Q^+(t)$, we have $\hat{q} = \ctrl(q, \tr_X(t))$
 and the right derivative $\dot{\tr_X}(t) = f_{\hat{q}}(\tr_X(t))$.
\end{itemize}

A trace is \emph{time divergent} if for all  $\Delta > 0$,
 $\mathsf{SwitchTimes}(\tr) \cap [0,\Delta]$ is a finite set.

\paragraph{Specification:} 
We want the continuous state $\vx$ to reach from initial (compact) set
$I \subset X$ to goal (compact) set $G \subset X$, while staying in 
safe (compact) sets $S \subset X$.
  
\begin{definition}[Reach While Stay]\label{Def:rws}
  The closed loop switched system $\Phi$ satisfies \RWS \ w.r.t
  $(I,G,S)$ iff for all traces $\tr$, $(\tr_X \in I) \implies
  (\tr_X \in S) \ \mathcal{U} \ (\tr_X \in G) $.
\end{definition}
To eliminate some technical arguments we assume $I \subseteq int(S)$.
The problem we study in this paper is synthesizing a controller that guarantees
\RWS \ w.r.t $(I,G,S)$ for the closed loop system. Also we are interested in
finding a big region $W \supseteq I$ s.t. the system satisfies \RWS w.r.t 
$(W,G,S)$.
\begin{problem}\label{prob:main}
  Given $S$, $G$ and $I \subseteq int(S)$ and a plant $\Psi$, 
  find a $\ctrl$ function and a region $W \supseteq I$ s.t. the closed loop
  switched system $\Phi$ satisfies \RWS \ w.r.t $(W,G,S)$.
\end{problem}

\section{Lyapunov Function for Switched Systems}
We recall Lyapunov and control Lyapunov functions for \RWS properties
of switched systems.
\begin{definition}~\label{def:lyap} 
A Lyapunov function for \RWS \ w.r.t $(I,G,S)$ is a
  continuous function $V: X \rightarrow \mathbb{R}^+$ iff
  there exists a constant $\beta$ s.t.
	\begin{compactenum}
		\item $(\forall \vx \in \partial S \setminus G) \ V(\vx) > \beta$
		\item $(\forall \vx \in I \setminus G) \ V(\vx) < \beta$
		\item $(\forall \tr, t''>t'\geq0) \  
		(\tr_X(t') \not \in int(G) \wedge V(\tr_X(t')) < \beta) \\ \implies
				V(\tr_X(t'')) < V(\tr_X(t'))$
	\end{compactenum}
\end{definition}

For a given trace $\tr$, let $\tr_V(t) = V(\tr_X(t))$. Because of the continuity
of $\tr_X$ and $V$, the third condition is equivalent to 
\begin{equation}
\begin{array}{l}
\exists \epsilon_Q, \ \forall\ \tr, \ (\forall t \geq 0) \ ( \tr_X(t) \not \in  G \wedge V(\tr_X(t)) < \beta) 
\\ \implies 
\dot{\tr_V}(t) = \frac{dV}{d\vx}
\dot{\tr_X}(t) < - \epsilon_Q
\end{array}
 \end{equation}
 This condition simply implies that the
value of Lyapunov function decreases through time.

\begin{definition}[Region for Lyapunov Function]
	Given a Lyapunov function $V$, let $\beta$ be the constant in Def. 
	~\ref{def:lyap}. The  associated region is defined as
$W :\ \{ \vx\in X\ |\ V(\vx) \leq \beta \} \cap S$.
\end{definition}

It is easy to show that (a) $W$ is a compact set, (b) $\vx \in \partial W \implies V(\vx) = \beta$, (c)
$I \setminus G \subseteq int(W)$ and (d) $W \setminus G \subseteq int(S)$. 
Ultimately we want to show $(\tr_X \in W) \implies (\tr_X \in W) \ U \ (\tr_X \in G)$
and since $W \subseteq int(S)$, the system satisfies \RWS \ w.r.t $(W,G,S)$.
Our overall strategy for controller synthesis is to synthesize a
(control) Lyapunov function. 
Such function can provide a control strategy as well as region $W$ such that
the closed loop system satisfies the specification.
However, while Lyapunov functions extend to proving stability for
switched/hybrid systems~\cite{liberzon2003switching}, care must be taken to
ensure that these techniques are not applied to systems with
\emph{time-convergent} traces. Defining the asymptotic behavior of
such traces as $t \rightarrow \infty$ is clearly not meaningful, when
the time $t$ never diverges.  It is possible for such trajectories to
``converge'' to a non-target state, even when a Lyapunov function
decreases.

Secondly, since our goal is to synthesize controllers,
time convergent behaviors represent physically unrealizable control strategies,
and must be avoided in our closed loop systems.
Therefore, it is quite essential that our controller synthesis
technique guarantee that the traces of the resulting closed loop
system are all time divergent.

\subsection{Control Lyapunov Function}

We focus, in this work, on finding polynomial control
Lyapunov functions for guaranteeing \RWS.

\begin{definition}[Control Lyapunov Function]
\label{def:clf}
A  control Lyapunov function (CLF) w.r.t. $(I,G,S)$ and a 
plant $\Psi$ is a polynomial function $V(\vx)$ iff there exist
a $\beta$ and $\epsilon_Q$ s.t.
\begin{equation}
    \label{Cond:negative-def} 
    \begin{array}{c}
    (\forall \vx \in \partial S \setminus G) \ V(\vx) > \beta \wedge \\
    (\forall \vx \in I \setminus G) \ V(\vx) < \beta \wedge \\
    (\forall \vx \in S \setminus G) \ (\exists q \in Q) \ \ 
        \dot{V}_q(\vx) = \frac{dV}{d\vx} f_q(\vx) < -\epsilon_Q
    \end{array}
\end{equation}        
\end{definition}

Given a control Lyapunov function $V$ w.r.t $(I,G,S)$, we define an
associated set of switching functions that define controllers, as
follows.

\begin{definition}[Switching Function for CLF]
Given a CLF $V$ and a function  $\ctrl: Q \times X \mapsto Q$, we
say that $\ctrl$ is compatible with $V$ iff 
for every state $\vx \in S \setminus G$ and mode $q$,
 the mode $\hat{q}: \ctrl(q, \vx)$ is such that $\dot{V}_{\hat{q}}(\vx) < 0$. 
\end{definition}

In other words, the controller at any state and any mode
chooses an input $q \in Q$ that makes the control Lyapunov function
decrease ``instantaneously''. 
Given a CLF, we wish to synthesize a controller that
establishes \RWS \ w.r.t $(W,G,S)$.
However, we cannot yet guarantee that the trajectories of the closed loop
system will all be time divergent.
As a result, we first tackle the problem of finding a switching function
that yields a minimum dwell time for each mode $q \in Q$. In other words,
whenever the controller switches into a mode $ q \in Q$, it must stay
in that mode for at least some time $\delta_m > 0$ before transitioning
to a different mode $\hat{q}$. 

\paragraph{Non-Zeno Switching Strategy:}
As usual, the goal of the
controller is to ensure that the CLF $V(\vx)$ decreases along
any trace. Suppose the current control mode is given by $q \in Q$.
Choosing a fixed constant  $\lambda > 1$, we define a switching function
as follows:
  \begin{equation}
	\label{eq:controller}
	 \ctrl(q, \vx) := \begin{cases} 
	\hat{q} & \mbox{if}\ 
		\left(\begin{array}{c} 
		\dot{V}_{q}(\vx) \geq - \frac{\epsilon_Q}{\lambda} \		\wedge \vx \in S \setminus G  \\
		\wedge \ \dot{V}_{\hat{q}}(\vx) < -\epsilon_{Q}
		\end{array} \right) \\ \\
	q &  \mbox{otherwise}
	\end{cases}
\end{equation}

The switching function above changes mode from $q$ to a new mode
$\hat{q}$ whenever the derivative of the CLF $\dot{V}_q$ is above a
threshold $- \frac{\epsilon_Q}{\lambda}$. The new mode it switches to
satisfies $\dot{V}_{\hat{q}} < - \epsilon_Q$. Such a mode is always
guaranteed to exist for $\vx \in S\setminus G$. Otherwise, the current
mode is retained. The main result shows that using the switching
function above, whenever the controller switches to mode $q$, there is
a fixed lower bound ($\delta_{m,q} > 0$) on the time before the
controller switches to another mode $\hat{q}$.

\begin{theorem}
\label{lem:delta-exists}
For each $q \in Q$, there exists a minimum dwell time $\delta_{m,q} > 0$ such that for any time $T > 0$ with $\tr_q(T) = q$, if (a) $\tr_X(T) \in S \setminus G$, (b) $\dot{V}_q(\tr_X(T)) < - \epsilon_Q$ and (c) $\tr_X(t) \in S \setminus
G$ for all $t \in [T,T+\delta_{m,q}]$, then $ (\forall t \in [T,
T+\delta_{m,q})) \ \dot{V}_{q}(\tr_X(t)) \leq
-\frac{\epsilon_Q}{\lambda}$.

As a corollary, the control mode does not change in time $[T,T+\delta_{m,q}]$: $\ctrl(\tr_q(t), \tr_X(t)) = \tr_q(T)$ for all $t \in [T,T+\delta_{m,q}]$.
\end{theorem} 

A closed form expression bound for $\delta_{m,q}$ is obtained in the
proof of Theorem~\ref{lem:delta-exists} as
$ \delta_{m,q} = \frac{(\lambda -1) \epsilon_Q } {\epsilon_1 \lambda}$,
wherein $\epsilon_1$ is a positive constant s.t. $\epsilon_1 \geq \min_{\vx \in S\setminus G} \ddot{V_q} 
$.
This is computed by
minimizing the polynomial second Lie derivative of $V$ w.r.t the dynamics
over the mode $q$ in the set $S \setminus G$. It can be solved conservatively
for instance through SOS programming~\cite{Parillo/2003/Semidefinite}.

 \begin{theorem}
 	\label{lem:system-reach-while-stay}
   Given compact sets $S$, $G$, $I \subseteq int(S)$, a plant $\Psi$
   and a CLF $V(\vx)$ w.r.t $(I,G,S)$ with associated region $W$,
   and a controller function $\ctrl$ that conforms to Equation ~\ref{eq:controller},
   the closed loop $\Phi$ satisfies the following properties.
   \begin{compactenum}
   \item System satisfies \RWS \ w.r.t $(W,G,S)$.
   \item All the traces of the closed loop system starting from $W$ 
   are time divergent before reaching $G$.
   \end{compactenum}
 \end{theorem} 

\paragraph{Discrete Controller:} Given $\delta_m$, a time-triggered discrete
controller computes the $\ctrl(q,\vx)$ function every $\delta_m$ time units. At
the start of each cycle (time $T$), the controller calculates $\dot{V}_q(\vx)$
and computes a mode it can safely switch to. Often many possible next modes may
exist, and the controller can use other performance criteria to choose the next
mode. Once chosen, this mode remains fixed until the beginning of next cycle
($[T+\tau]$). Theorem~\ref{lem:delta-exists} guarantees that value of CLF
decreases during each cycle and therefore, we make progress towards our goal
$G$.

\section{Synthesizing CLFs}
The described solution in previous section reduces Problem~\ref{prob:main} to 
finding a CLF. In this section, we focus
on the problem of searching a CLF. First, we introduce a general
CEGIS framework for solving $\exists\forall$ formula for real arithmetics.

The general problem we wish to solve has the following form:
\begin{equation} 
(\exists \vc \in C)
\begin{cases}
	\bigvee_j \bigwedge_k G_{j,k}(\vc) < 0 \\
		(\forall \vx \in R_1) \bigvee_k F_{1,k}(\vx, \vc) < 0 \\
		(\forall \vx \in R_2) \bigvee_k F_{2,k}(\vx, \vc) < 0 \\
		\cdots \\
		(\forall \vx \in R_m) \bigvee_k F_{m,k}(\vx, \vc) < 0
\end{cases}		
\label{eq:cegis-general}
\end{equation}
where $R_j$ is a fixed compact region and $F_{j,k}$ are functions polynomial in $\vx$
and linear in $\vc$. Also $G_{j,k}$ are functions linear in $\vc$. 
The CEGIS algorithm was first introduced to tackle $\exists\forall$
 constraints such as ~\eqref{eq:cegis-general} by Solar-Lezama
 et al.~\cite{solar2006combinatorial,SolarLezema/08/Program}. The
 basic idea is to maintain two sets:
\begin{compactenum}
\item A finite set of \emph{witnesses}: $X_i : \{ \vx_1, \ldots, \vx_{n_i}\} \subseteq X$,
 namely, the $X$-space.
\item A subset $C_i \subseteq C$, namely, the $C$-space.
\end{compactenum}

The $C$-space represents the set of candidates which are to be
examined by our procedure while the $X$-space represents \emph{test
  points} over which a candidate is tested. The $i^{th}$ 
iteration involves  the following steps:
\begin{compactenum}[Step 1)]
\item\label{Step:sat-check} Choose an arbitrary $\vc_i \in C_i$ to get candidate
solution for Eq.~\eqref{eq:cegis-general}.
\item\label{Step:clf-check} Check if Eq.~\eqref{eq:cegis-general}
holds for $\vc_i$.
\begin{compactenum}[(a)]
\item If Eq. ~\eqref{eq:cegis-general} holds, procedure terminates 
immediately.
\item Otherwise, a point $\vx_i$ is obtained at which Eq.~\eqref{eq:cegis-general} 
fails. $\vx_i$ is added to the set of test points ($X_{i+1} : X_i \cup \{x_i\}$).
\end{compactenum}
\item\label{Step:cspace-ref} $C$-Space is refined by removing all candidates 
which fail at $\vx_i$ by not satisfying~\eqref{eq:cegis-general}
\end{compactenum}

The procedure terminates successfully if a solution is
found. Alternatively, if $C_j = \emptyset$ then it terminates without
finding a solution. Finally, the procedure may run forever.

\paragraph{Representing the $C$-space:} Each set $C_i$ is represented
using a linear arithmetic formula $\psi_i[\vc]$
 such that $C_i:\ \{ \vc \ |\ \psi_i[\vc]\
\mbox{holds}\}$. The initial formula $C_0$ is simply
$\psi_0: \bigvee_j \bigwedge_k G_{j,k}(\vc) < 0$.  
Step~\ref{Step:sat-check} is implemented using a SMT solver to check if $\psi_i$ is
satisfiable and obtain a candidate $\vc_i$ as a solution to $\psi_i$~\cite{DeMoura+Bjorner/08/Z3}.
Likewise, step~\ref{Step:cspace-ref} is implemented by augmenting $\psi_i$ to yield
a formula 
\begin{equation}	\label{eq:clf-linear-constr}
	\begin{array}{c}
		\psi_{i+1}: \psi_i\ \land\ \bigwedge_j \psi_{i,j}	\end{array}
\end{equation}
where $\psi_{i,j}$ is \emph{True} if $\vx_i \not \in R_j$ and 
$\bigvee_k F_{j,k}(\vx_i, \vc) < 0$ otherwise.

\paragraph{Finding Witnesses:} On the other hand, finding witnesses requires us to 
check the satisfiability of a non-linear constraints obtained by negating~\eqref{eq:cegis-general}:
\begin{equation}\label{Eq:cegis-nonlinear-constr}
	\begin{array}{l}
\vx \in R_1 \land \bigwedge_k F_{1,k}(\vx, \vc_i) \geq 0\\ \cdots \\
\vx \in R_m \land \bigwedge_k F_{m,k}(\vx, \vc_i) \geq 0
	\end{array}
\end{equation} If yes, a
witness $\vx_i$ is obtained at which the current candidate $\vc_i$ fails to satisfy
Eq.~\ref{eq:cegis-general}. Otherwise, we conclude problem is solved. However,
solving this constraint requires a nonlinear arithmetic solver that is capable
of finding witnesses. Relaxations such as SOS programming can
be used to check whether 
the formula~\eqref{Eq:cegis-nonlinear-constr} is
unsatisfiable~\cite{Parillo/2003/Semidefinite}, but failing this, they
do not provide useful witnesses to generate future candidates.
Therefore, we resort to a promising approach for
checking~\eqref{Eq:cegis-nonlinear-constr} implemented in the tool
dReal~\cite{DBLP:conf/cade/GaoKC13}. dReal checks if the formula is
unsatisfiable, and if it reports UNSAT, we conclude that the
constraints~\eqref{Eq:cegis-nonlinear-constr} are indeed unsat. Otherwise, it
reports that a $\delta$-perturbation is satisfiable, and provides us a witness.
Therefore, using dReal, we run the risk of obtaining additional possibly
spurious witnesses, and not recognizing if a solution has already been found.
However, the resulting procedure will not yield a wrong solution.

\subsection{Searching for CLF}
Given a plant $\Psi$, and regions $S$, $G$ and $I$,
We fix a \emph{template} polynomial form $V(\vx,\va)$
parameterized by variables in $\va \in A$ as the desired CLF. The
space $A$ is a compact set chosen as a hyper-rectangle
limiting each $a_i \in [L_i,U_i]$. Formally, we wish to find
$\vc = (\va, \epsilon_Q, \beta) \in C : \reals^m \times \reals \times \reals$ 
that satisfies the conditions
 in Def.~\ref{def:clf}:
 \begin{equation}\label{eq:chatter-free-template}
 (\exists\ \vc \in C)
 \begin{cases}
 (\epsilon_Q > 0 \wedge \bigwedge_k a_k > L_k \wedge a_k < U_k) \\
 (\forall \vx \in \partial S \setminus G) \ V(\vx, \va) > \beta \land \\
 (\forall \vx \in I \setminus G) \ V(\vx, \va) < \beta \land \\ 
 (\forall \vx \in S \setminus G) \left( \begin{array}{c}
 \mathop{\bigvee}\limits_{q \in Q} \dot{V}_q(\vx, \va) <  -\epsilon_Q
\end{array}\right)
 \end{cases}
 \end{equation}
 
Since the form $V(\vx,\va)$ is linear over the parameters in $\va$, the above
Equation is a typical case that can be solved by the CEGIS framework described 
above.

\subsection{Adopting CEGIS to Real Arithmetics}
We now briefly discuss the termination of the CEGIS procedure. We noted
that termination is possible if a solution of the desired form in~\eqref{eq:cegis-general}
exists, or alternatively, the $C$-space is exhausted. However, neither
situation may result and the algorithm may run forever. In this section,
we provide a simple strengthening of Eq.~\eqref{eq:clf-linear-constr} that
guarantees termination.
We strengthen~\eqref{eq:clf-linear-constr} as follows (when $\vx_i \in R_j$):
 \begin{equation}\label{eq:relaxed-termination}
 \psi_{i,j}: \vee_k F_{j,k}(\vx_i, \vc) < - \epsilon_{T_j}
 \end{equation}
wherein $\epsilon_{T_j} > 0$ are positive constants. 
The two constraints are identical when $\epsilon_{T_j} = 0$. Let
 $\vc_i$ be a candidate examined at the $i^{th}$ iteration of the
 CEGIS procedure modified to use Eq.~\eqref{eq:relaxed-termination}.
 Suppose there exists a counter example $\vx_i$ corresponding to $\vc_i$.
 We compute a new refined $C$-space  $C_{i+1}$. It is easily shown the $\vc_i \not \in C_{i+1}$.
Furthermore, by using~\eqref{eq:relaxed-termination}, we obtain the following result
that any candidate in a $\eta$-ball around $\vc_i$ is also eliminated.
\begin{theorem}
   \label{lem:eta-exists}
  If the CEGIS procedure were modified using
  Eq.~\eqref{eq:relaxed-termination} with a given $\epsilon_{T_j} > 0$,
  then there exists a constant $\eta > 0$ such that at each
  iteration $i$, $ \mathcal{B}_{\eta}(\vc_i) \cap C_{i+1} =
  \emptyset$.
\end{theorem}

As a result, starting from a initial set $C_0$, given $C_0$ is a compact set,
we note that employing the stronger rule~\eqref{eq:relaxed-termination} guarantees
that at each step, an $\eta$-ball around the current solution is also
removed. Thus, either a CLF is found or
the $C$-space is empty in finitely many iterations.
If we exhaust the
$C$-space for a given values of $\epsilon_{T_j}$ it is possible to repeat the
search by halving $\epsilon_{T_j}$ to alleviate against the loss of possible
solutions due to the strengthening of Eq.~\eqref{eq:clf-linear-constr}
by ~\eqref{eq:relaxed-termination}.

\paragraph{Faster Termination}

A first cut application of the CEGIS approach, presented thus far,
resulted in a prohibitively large number of witnesses, failing on most
of our benchmarks. This happens because the witnesses and candidate
functions returned by the SMT solvers are similar (close in term of
Euclidean distance).  We discuss a heuristic to select witnesses
$\vx_i$ at each step of the CEGIS procedure, that led to the successful
implementation of the overall procedure.

Given a current candidate $\vc_i$,  we may split the search for a witness into $m$ 
parts: find a witness that violates the 
$\bigvee_k F_{j,k}(\vx, \vc_i) < 0$  
(for each $1 \leq j \leq m$).
We will search for a counterexample that produces the ``most-egregious'' violation
of the constraints possible. Therefore, we wish to maximize 
$\min_k F_{j,k}(\vx, \vc_i)$.
However, solvers such as dReal currently lack the ability to optimize. Therefore,
we simply fix a constant $\gamma \geq 0$ and search for  $\vx_i$
 satisfying
$\bigwedge_k F_{j,k}(\vx, \vc_i) - \gamma \geq 0$.
A larger $\gamma$ leads to a more ``egregious'' violation and a larger set
of candidates ruled out in the $C$-space and it is less likely to find a 
candidate that is similar to the previously selected candidate. 
The parameter $\gamma$ itself
is iteratively reduced to find a witness or conclude that no witness exists
when $\gamma = 0 $.

Also, the method can considerably get improved by seeding with an initial
 set of points $X_0$.

  \subsection{Complexity and Incompleteness}
There are many sources of incompleteness: (a) 
The polynomial template on the CLF with a maximum  degree;
(b) The use of $\epsilon_{T_j}$ in
 Eq.~\ref{eq:relaxed-termination}; and finally (c) the use of a
 $\delta$-satisfiability solver for nonlinear constraints. However, it
 is possible to reduce this incompleteness by making $\delta$ smaller.

In terms of complexity, solving linear arithmetic constraints and quantifier
free nonlinear constraints are well-known to be NP-hard. In addition, while it
is guaranteed that there will be a finite number of iterations in the CEGIS
procedure, this number can be really large. Though we provided some heuristics
to decrease the number of iterations, the worst case can be in the order of
$O(d^m)$, where $m$ is the number of unknown coefficients in the template and
$d$ is a function of $L_i$, $U_i$ and $\eta$ in Theorem~\ref{lem:eta-exists}.

\section{Evaluation and Discussion}



Our approach was implemented as a Python script that wraps around the
Z3~\cite{DeMoura+Bjorner/08/Z3} and
dReal~\cite{DBLP:conf/cade/GaoKC13} solvers.  The inputs to our
procedure include a description of the plant model, the set $S$ (taken
to be a box), the sets $G,I$ are provided as balls of radius $\sigma_G$
and $\sigma_{I}$, respectively. In addition, we assume $\epsilon_Q$ is given
as input. Our approach requires parameters
$\epsilon_{T_1} = \epsilon_{T_2}$ corresponding to the first two
inequalities in Eq.~\eqref{eq:chatter-free-template}, and
$\epsilon_{T_3}$ for third one. The choice of these parameters is
currently manual, but we are investigating automatic selection
heuristics as part of our ongoing work. Finally, we assume a quadratic
CLF template for all benchmarks, with the template
coefficients belonging to a compact set  $A: \prod_i [L_i, U_i]$, that can be
specified by the user.

\paragraph{Benchmarks} We collected $20$ 
benchmark instances that are used in our evaluation. These benchmarks
are taken from many sources and adapted to produce problem instances
for our evaluation
~\cite{liberzon1999basic,perruquetti1996lyapunov,saat2011nonlinear,greco2005stability,Mazo+Others/2010/PESSOA,PESSOA:Website,nilssonincremental,mouelhi2013cosyma,gol2012finite,URL:2014:Online,pettersson2001stabilization,zhang2009exponential,faubourg1999design}.
We manually formulated a \RWS specification where one was not
available. Finally, our approach does not yet consider disturbances
--- benchmarks with disturbances were modified by setting to nominal
values.
\emph{A detailed description of each
benchmark can be found in the appendix.} The results  are summarized in Table~\ref{tab:result}.

\begin{table*}[t]
\caption{Results of running our implementation on the
  benchmark suite}
\label{tab:result}

\textbf{Legend}: $n$: \# state variables, $|Q|$: \# modes,
  , $\delta$: dReal precision,
  $itr$ : \# iterations, time: total computation time, Z3 T: time taken by Z3,
  dReal T: time taken by dReal, OM: Out of Memory, \tick: Proper Radial CLF Found,
  \crossMark: Failed. All timings are in seconds.
 
\begin{center}
{
\begin{tabular}{||l|l|l|l||l|l|l||l|r|r|r|l||l|r|l||}
\hline
\multicolumn{4}{||c||}{Problem} & \multicolumn{3}{c||}{Parameters}
& \multicolumn{5}{c||}{Results} & \multicolumn{3}{c||}{Other tools}
\tabularnewline \hline
ID & $n$ & $|Q|$ & $\epsilon_Q$ & $\epsilon_{T_1}$ & $\epsilon_{T_3}$ & $\delta$
& $itr$ & z3 T & dReal T & Tot. Time & Status & Tool & Tool Time & Rem.
\tabularnewline \hline
1 & 2 & 2 & 0.01 & 0.1 & 0.01 & $10^{-5}$ &
15 & 0.4 & 4.6 & 5.3 & \tick & \multicolumn{3}{c||}{ -- NA --  }
\tabularnewline \hline
2 & 2 & 2 & 0.0001 & 0.1 & 0.1 & $10^{-6}$ &
15 & 0.5 & 5.6 & 6.6 & \tick & \multicolumn{3}{c||}{ -- NA --  }
\tabularnewline \hline
3 & 2 & 2 & 0.001 & 0.1 & 0.1 & $10^{-5}$&
7 & 0.0 & 2.3 & 2.5 & \tick & \multicolumn{3}{c||}{ -- NA --  }
\tabularnewline \hline
4 & 2 & 5 & 0.0001 & 0.1 & 0.0001 & $10^{-6}$&
1 & 0.0 & 0.8 & 0.8 & \tick & \multicolumn{3}{c||}{ -- NA --  }
\tabularnewline \hline
5 & 2 & 2 & 0.01 & 0.1 & 0.01 & $10^{-6}$&
3 & 0.0 & 3.4 & 3.6& \tick & PESSOA & 42.8 & (\textsc{r1})
\tabularnewline \hline
6 & 2 & 3 & 0.05 & 0.1 & 0.05 & $10^{-5}$ &
13 & 0.1 & 49.2 & 50.0 & \tick & PESSOA & 6881.1 & (\textsc{r1})
\tabularnewline \hline
7 & 2 & 2 & 0.001 & 0.1 & 0.001 & $10^{-4}$ &
6 & 0.1 & 1.6 & 2.0 & \tick & \multicolumn{3}{c||}{ -- NA --  }
\tabularnewline \hline
8 & 2 & 2 & 0.1 & 0.1 & 0.1 & $10^{-7}$ &
6 & 0.1 & 3.6 & 4.0 & \tick & CoSyMA  & 3.2  & (\textsc{r2})
\tabularnewline \hline
9 & 3 & 4 & 0.001 & 0.1 & 0.001 & $10^{-4}$ &
1 & 0.0 & 2.8 & 2.8 & \tick & CoSyMA  & 1.8  & (\textsc{r1})
\tabularnewline \hline
10 & 3 & 4 & 0.05 & 0.2 & 0.05 & $10^{-4}$ &
8 & 4.4 & 80.8 & 86.2 & \tick & \multicolumn{3}{c||}{ -- NA --  }
\tabularnewline \hline
11 & 3 & 3 & 0.0001 & 0.1 & 0.01 & $10^{-5}$ &
15 & 25.3 & 59.6 & 86.3 & \tick & \multicolumn{3}{c||}{ -- NA --  }
\tabularnewline \hline
12 & 3 & 5 & 0.0001 & 0.1 & 0.1 & $10^{-5}$ &
8 & 8.0 & 41.4 & 50.4 &  \tick & \multicolumn{3}{c||}{ -- NA --  }
\tabularnewline \hline
13 & 3 & 2 & 0.001 & 0.5 & 0.5 & $10^{-5}$ &
17 & 61.7 & 116.1 & 179.8 &  \tick & \multicolumn{3}{c||}{ -- NA --  }
\tabularnewline \hline
14 & 3 & 2 & 1.0 & 0.1 & 10.0 & $10^{-5}$ &
36 & 48.1 & 57.3 & 108.4 & \tick & ~\cite{nilssonincremental} & 5319.5 & see Fig.~\ref{fig:region-comparison}
\tabularnewline \hline
15 & 4 & 5 & 0.001 & 0.1 & 0.001 & $10^{-4}$ &
1 & 0.0 & 27.8 & 27.8 & \tick & CoSyMA  & OM (494.0) & (\textsc{r3})
\tabularnewline \hline
16 & 4 & 2 & 0.0001 & 0.1 & 0.01 & $10^{-5}$ &
4 & \multicolumn{3}{c|}{ -- \textgreater 1hr --  } & \crossMark & \multicolumn{3}{c||}{ --  NA --  }
\tabularnewline \hline
17 & 4 & 2 & 0.001 & 0.1 & 0.1 & $10^{-6}$ &
4 & \multicolumn{3}{c|}{ -- \textgreater 1hr --  }  & \crossMark & \multicolumn{3}{c||}{ --  NA --  }
\tabularnewline \hline
18 & 5 & 6 & 0.001 & 0.1 & 0.001 & $10^{-4}$ &
1 & 0.0 & 649.7 & 650.0 & \tick & CoSyMA & OM (571.0) & (\textsc{r3})
\tabularnewline \hline
19 & 6 & 4 & 0.001 & 0.1 & 0.001 & $10^{-4}$ &
2 & 0.5 & 2994.0 & 2995.6 & \tick & CoSyMA & OM & 
\tabularnewline \hline
20 & 9 & 4 & 0.001 & 0.1 & 0.001 & $10^{-4}$ &
1 & \multicolumn{3}{c|}{ -- \textgreater 1hr --  } & \crossMark & CoSyMA & OM & 
\tabularnewline \hline

\end{tabular}\\
}

\end{center}

(\textsc{r1}): Parameters as reported in the related works~\cite{PESSOA:Website,mouelhi2013cosyma}.
 (\textsc{r2}): Parameters: $N = 2$, $\tau = 0.1$, $\eta = 0.008$.
Controllability Ratio 47.2\% ~\cite{mouelhi2013cosyma}
(\textsc{r3}): Couldn't reproduced (OM). Timings as reported in
~\cite{mouelhi2013cosyma}.
\vspace*{-0.3cm}
\end{table*}

On the positive side, our approach finds a \emph{CLF} for $18$ out of
the $20$ benchmark instances. Our technique was successful on some
benchmarks with upto $6$ state variables. However, our approach timed
out on $2$ of the larger instances: the nonlinear solver dReal was
responsible for the timeout in each case.

\paragraph{Preliminary Comparison}
We also considered a \emph{preliminary comparison} with three
implementations: the PESSOA tool by Mazo et al.
~\cite{Mazo+Others/2010/PESSOA}, the CoSyMa tool by Camara et
al.~\cite{camara2011synthesis,mouelhi2013cosyma} and the prototype
corresponding to the recent work by Nilsson et
al.~\cite{nilssonincremental}. The implementations for the other
related approaches could not be obtained.

\begin{figure}[t]
\begin{center}
\includegraphics[width=0.45\textwidth]%
	{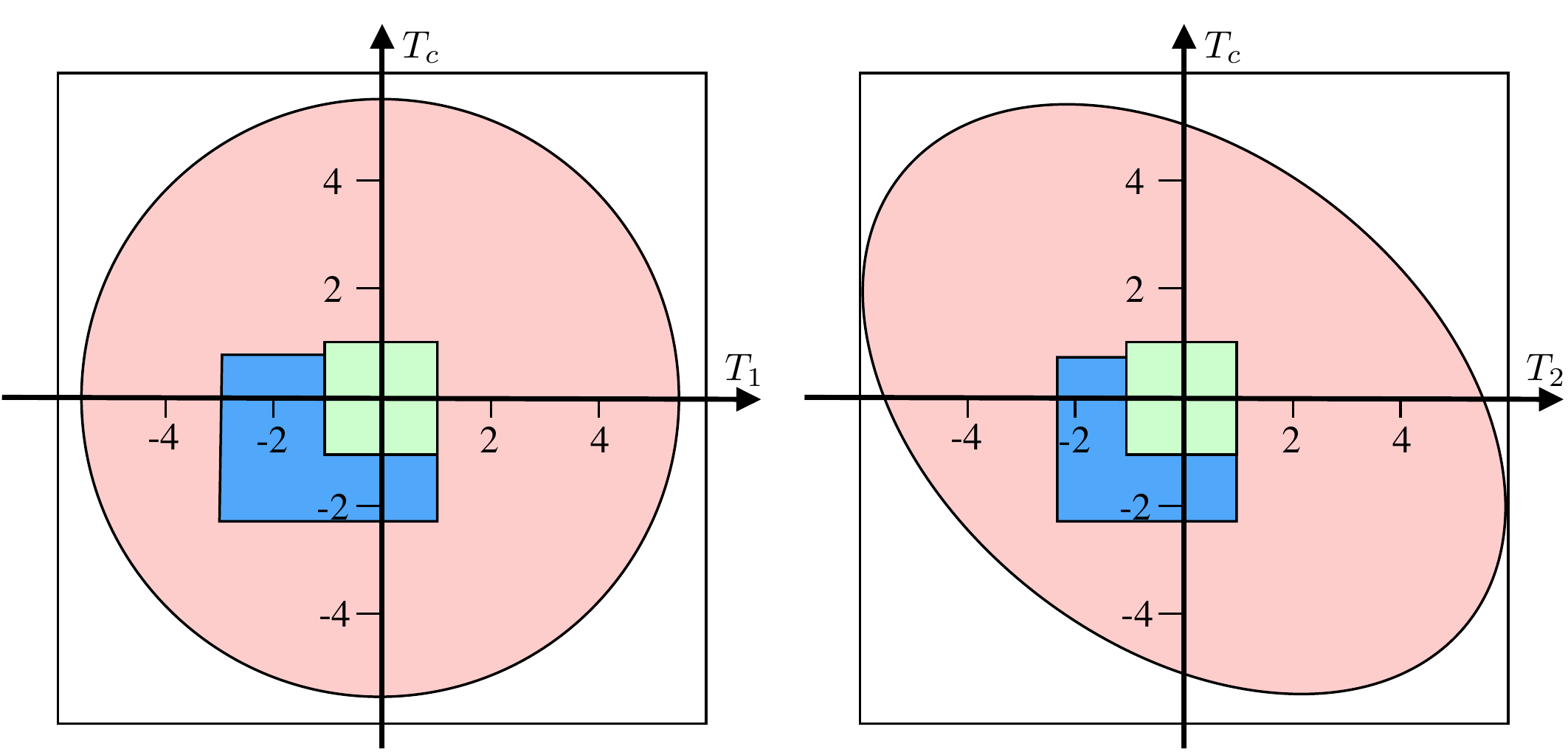}
\caption{Region $W$ (red) is for our method
and the winning region after 350 iterations (blue)
 for the approach of Nilsson et al.~\cite{nilssonincremental}.}
\label{fig:region-comparison}
\end{center}
\vspace*{-0.5cm}
\end{figure}

Unfortunately, just 8 out of the 20 benchmarks could be successfully
compared. Reasons included the lack of support for some required
features, implementation issues and the lack of proper
documentation. Therefore, the comparisons are restricted to the $8$
cases that either (a) ran successfully on our machines, or (b)
instances whose results/running time were reported in the
corresponding references.  Table~\ref{tab:result} shows a comparison
between our method and these methods using examples chosen from
referenced papers.  We found that our technique is \emph{faster} on
almost all benchmarks compared, even while allowing for the
differences in the implementation platforms. We attribute this to many
reasons: (A) Our approach is currently specialized to \RWS, whereas
other approaches consider generic LTL properties. Nevertheless, all
comparisons involved solving \RWS problems. (b) Building a finite
abstraction is very expensive even for systems with $2$ or $3$
dimensions, and this takes a majority of the time in these
benchmarks. Our approach does not construct abstractions
explicitly. Finally, Fig~\ref{fig:region-comparison} compares the
regions $W$ (an ellipsoid) obtained for system with ID 18 against the
winning region for the \RWS \ property using the CEGAR-based
approach~\cite{nilssonincremental}.


\begin{figure}[t]
\begin{center}
\includegraphics[width=0.4\textwidth]%
	{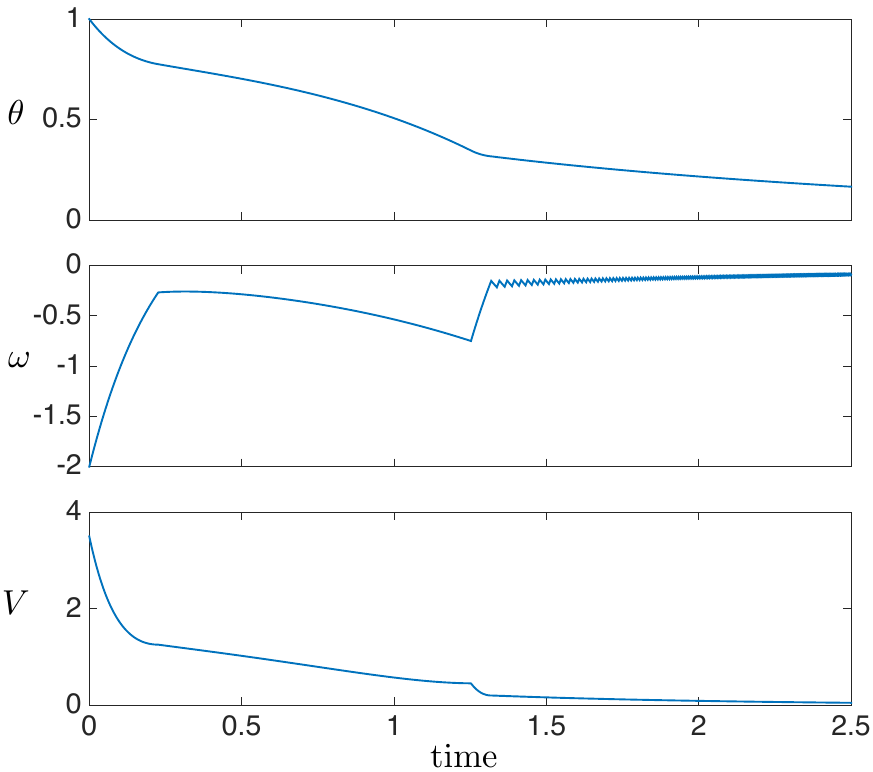}
\caption{Simulation of execution trace of the Inverted Pendulum example }
\label{fig:simulation}
\end{center}
\vspace*{-0.7cm}
\end{figure}

\paragraph{Simulation:} The dynamics of the  inverted pendulum on a 
cart ~\cite{PESSOA:Website} example (ID 7) are
given as $ \dot{\theta} =\omega,\ \ \ \dot{\omega}
=\frac{g}{l}sin(\theta)-\frac{h}{ml^2}\omega+\frac{1}{ml}cos(\theta)u\,,
$ where $g$, $h$, $l$ and $m$ are constants. We used Taylor
expansion to approximate the trigonometric function, and the input
$u$ is discretized to be in set $\{-30, 30\}$. Considering region $S =
[-1.5, 1.5]\times[-4, 4]$, $\sigma_I = 0.5$, $\sigma_G = {0.2}$,
$\epsilon_Q = 0.05$ and parameters $\epsilon_{T_1} = 0.1$, $\epsilon_{T_3} =
0.05$ and $\delta = 10^{-5}$, we find the CLF $V([\theta \ \omega]^T) =
0.65625y^2 + 0.69043xy + 2.2539x^2$.  
The underapproximate minimal dwell time $\tau =
0.0002s$. Figure~\ref{fig:simulation} shows a simulation with initial
state $[\theta, \omega] = [1 \ -2]$.

\section{Conclusion}

We have demonstrated a CEGIS procedure for synthesizing CLFs for
switched systems to ensure \RWS and satisfy a minimal dwell time
requirement.  We have demonstrated some preliminary evidence of the
applicability of our approach to many examples.  Moving forward, we
are exploring the use of relaxations such as sum-of-squares (SOS)
programming and Bernstein polynomials, while ascribing witnesses when
the formula turns out to be satisfiable. 
\bibliographystyle{IEEEtran}









%








\newpage
\appendices

\section{Proofs}

\paragraph{Proof of  Theorem~\ref{lem:delta-exists}}
For each $q \in Q$, there exists a minimum dwell time $\delta_{m,q} > 0$ such that
 for any time $T > 0$ with $\tr_q(T) = q$, if (a) $\tr_X(T) \in S \setminus G$, 
 (b) $\dot{V}_q(\tr_X(T)) < - \epsilon_Q$ and (c) $\tr_X(t) \in S \setminus
G$ for all $t \in [T,T+\delta_{m,q}]$, then $ (\forall t \in [T,
T+\delta_{m,q})) \ \dot{V}_{q}(\tr_X(t)) \leq
-\frac{\epsilon_Q}{\lambda}$.

A constructive proof is provided here. 
Assume at time $T$ s.t. 
\begin{equation}
    \label{eq:vq-leq-neg-phi}
\dot{V}_{q}(\tr_X(T)) < - \epsilon_{Q}
\end{equation}
and let $T + \delta$ ($\delta > 0$) be the minimum time where 
\[\dot{V}_{q}(\tr_X(T+\delta)) \geq -\frac{\epsilon_Q}{\lambda}\]
if \[ (\forall t \in [T , T + \delta ])\  
\tr_Q(t) = q \wedge \tr_X(t) \in S \setminus G\]
Let  $\ddot{V}_q : X \rightarrow \mathbb{R}$
 be 
 \begin{align*}
 \ddot{V}_q = \frac{d\dot{V}_q}{d\vx} f_q(\vx) \\
 \end{align*}

Since $S \setminus G$ is a bounded set and $\ddot{V}_q$ is a polynomial, 
there exist $\epsilon_1 > 0$
s.t.
\begin{align}
(\forall \ \vx \in S \setminus G) \ \ddot{V}_q (\vx) & \leq \epsilon_1 
\label{eq:ddV-leq-epsilon-phi}
\end{align}
Also
\begin{align*}
     \dot{V}_{q}(\tr_X(T+\delta)) & = \dot{V}_{q}(\tr_X(T)) + \int_{T}^{T+\delta} 
    \ddot{V}_{q}(\tr_X(t)) dt \\
\overset{Equation~\ref{eq:ddV-leq-epsilon-phi}}{\implies}    &
  \leq  \dot{V}_{q}(\tr_X(T)) + \epsilon_1 \delta
\end{align*}
Then, we can conclude 
\begin{align*}
   - \frac{\epsilon_Q}{\lambda} < -\epsilon_Q + \epsilon_1 \delta
   \implies 
   \frac{(\lambda-1) \epsilon_Q}{\epsilon_1 \lambda} \leq \delta
\end{align*}

The above arguments suggest that if
 $\dot{V}_{q}(\tr_X(T)) < - \epsilon_Q$ for a mode $q$,
then without switching there exists $\delta_{m,q}>0$ ($\delta$ in above argument)
s.t. 
\[
(\forall t \in [T, T+\delta_{m,q}]) \ \dot{V}_{q}(\tr_X(t)) \leq - \frac{\epsilon_Q}{\lambda})
\]

\paragraph{Proof of  Theorem~\ref{lem:system-reach-while-stay}}
   Given compact sets $S$, $G$, $I \subseteq int(S)$, a plant $\Psi$
   and a CLF $V(\vx)$ w.r.t $(I,G,S)$ with associated region $W$,
   and a controller function $\ctrl$ that conforms to Equation ~\ref{eq:controller},
   the closed loop $\Phi$ satisfies the following properties.
   \begin{compactenum}
   \item System satisfies \RWS \ w.r.t $(W,G,S)$.
   \item All the traces of the closed loop system starting from $W$ 
   are time divergent before reaching $G$.
   \end{compactenum}

First, we show that $I \setminus G \subseteq int(W)$ and $W \setminus G \subseteq int(S)$.
Remember $W : \{\vx | V(\vx) \leq \beta\} \cap S$.
By definition $W \setminus G \subseteq S$
and $(\forall \vx \in \partial S) \ V(\vx) > \beta$. Therefore, $W \subseteq 
int(S)$. Also, $(\forall \vx \in I \setminus G) \ V(\vx) < \beta$ and consequently
 $I \setminus G \subseteq int(W)$. 

    According to Theorem~\ref{lem:delta-exists}, 
    whenever there is a switch to a mode $q$, then the output
    of the controller remains $q$ for at least $\delta_{m,q}$ time unit
    unless trace reaches $G$ or leaves $S$.
    
    Now we show that $(\tr_X \in W) \implies (\tr_X \in W)  \ U \  (\tr_X \in G)$.
    
    Assume that $\tr_X(0) \in (W \setminus G)$. $\tr_X$ can not leave $S$ before
    reaching $\partial W$ or $G$. Let $T \geq 0$ be the first time that $\tr_X(T)$
    reaches $\partial W$ before reaching $G$. Then $\tr_V(T) = \beta$. 
    Since $\tr_V(0) \leq \beta$ and 
    $(\forall 0 \leq t < T) \ \dot{\tr_V}(t) < 0$, $T$ has to be zero.
    At start ($t = 0$), since 
    $\tr_X(0) \in (W \setminus G)$, then the controller
    choose a mode $q$ such that $\dot{\tr_V}(0) < 0$ and therefore, the trace
    leaves the boundary of $W$ and the trace goes to the interior of
    $W$ ($int(W)$).
    
    If
    $(\forall t > 0) \ \tr_X(t) \in W \setminus G$, by the construction of the
    controller, we can conclude time diverges. Also since $\dot{\tr_V}(t) < 
    -\frac{\epsilon_Q}{\lambda}$ for all times, $\tr_V$ decreases to infinity. 
    However, the value of $V$ is bounded on bounded set $W \setminus G$.
    Therefore, $\tr_X$ can not remain in $W \setminus G$ and can not reach 
    the boundary of $W$. The only possible outcome for the trace is to reach 
    $G$. System satisfies \RWS \ w.r.t $(W,G,W)$ and therefore it satisfies \RWS
    \ w.r.t $(W,G,S)$.

\paragraph{Proof of  Theorem~\ref{lem:eta-exists}}
  If the CEGIS procedure were modified using
  Eq.~\eqref{eq:relaxed-termination} with a given $\epsilon_{T_j} > 0$,
  then there exists a constant $\eta > 0$ such that at each
  iteration $i$, $ \mathcal{B}_{\eta}(\vc_i) \cap C_{i+1} =
  \emptyset$.

Given a counter example $\vx_i \in R_j$ for $\vc_i$, the following is true
 \begin{align}
  \label{eq:counter-example-result}
  \wedge_k \ F_{j,k}(\vx_i, \vc_i) \geq 0
 \end{align}
 and the added restriction on \emph{C-Space} for next iteration is
 \begin{align}
  \label{eq:next-iteration-restrictions}
  \vee_k \ F_{j,k}(\vx_i, \vc_i) < - \epsilon_{T_j}
 \end{align}
 Let $F'_{j,k}$ be polynomial $F_{j,k}$ without monomials which do not have variables
 in $\vc$.
 Since $F'_{j,k}$ is a polynomial in $\vx$ and linear in $\vc$,  and $R_j$ is a 
 compact set, there exists $M_{j,k} > 0$ s.t. $(\forall \vc \in C_0) \ (\forall \vx \in R_j) \
 F'_{j,k}(\vx, \vc) \geq - M_{j,k}$ and as a result, there exists a $\eta > 0$ s.t. 
 \begin{equation}
 	\label{eq:eta-ball-inequality}
 	(\forall \vc \in \mathcal{B}_{\eta}(\vzero)) \ 
 (\forall \vx \in R_j) (\forall k) \ F'_{j,k}(\vx, \vc) \geq - \epsilon_{T_j}
 \end{equation}

  Also,
 \begin{align*}
  (\forall \vc \in \mathcal{B}_{\eta}(\vc_i)) \ (\forall k)& \\
  F_{j,k}(\vx_i, \vc) & = F_{j,k}(\vx_i, \vc_i) + F'_{j,k}(\vx_i, 
  \vc_\eta)
 \end{align*}
 where $\vc_\eta \in \mathcal{B}_{\eta}(\vzero)$. Therefore, by Equations ~\ref{eq:counter-example-result}
 and ~\ref{eq:eta-ball-inequality}
 \[(\forall \vc \in \mathcal{B}_{\eta}(\vc_i)) \ 
 \wedge_k F_{j,k}(\vx_i, \vc) \geq  - \epsilon_{T_j}\]
 and by comparing this to Equation~\ref{eq:next-iteration-restrictions}
  it is easy to see that $\vx_i$ is a counter example for $\vc \in \mathcal{B}_{\eta}(\vc_i)$.
and $\mathcal{B}_{\eta}(\vc_i) \cap C_{i+1} = \emptyset$ .

\section{Benchmark}

The benchmark used in the experiments are examples adopted from
literature. We consider each of these systems as a switched system with \RWS \ as
the specification.

\begin{system} 
\label{sys:harmonic}
This system is adopted from ~\cite{liberzon1999basic}.
There are two continuous variables $x$ and $y$ and the dynamics are
\begin{align*}
    \dot{x} & = y \\
    \dot{y} & = - x + u
\end{align*}
, where $u \in \{-1, 1\}$. $S = [-1 \ \ 1]^2$, $\sigma_{I} = 0.5$ 
and $\sigma_G = 0.1$.
\end{system}

\begin{system}
\label{sys:sliding-motion-2}
This system is adopted from ~\cite{perruquetti1996lyapunov}.
There are two continuous variables $x$ and $y$ and the dynamics are
\begin{align*}
    \dot{x} & = u \\
    \dot{y} & = y^2x
\end{align*}
, where $u \in \{-4, 4\}$.  And region $S = [-1 \ \ 1]^2$, $\sigma_{I} = 0.5$ 
and $\sigma_G = 0.1$.
\end{system}

\begin{system}
\label{sys:ETH-example-1}
This system is adopted from ~\cite{URL:2014:Online}.
There are two continuous variables $x$ and $y$ and the dynamics are
\begin{align*}
    \dot{x} & = y - x^3\\
    \dot{y} & = u
\end{align*}
, where $u \in \{-1, 1\}$. The region of interest is $S = [-1 \ \ 1]^2$, 
$\sigma_{I} = 0.5$ and $\sigma_G = 0.05$.
\end{system}

\begin{system}
\label{sys:linear-ss-1}
This system is a switched system
adopted from ~\cite{greco2005stability}.
There are two continuous variables $x$ and $y$ and
5 modes ($q_1,..., q_5$) the dynamics of each mode 
is described below
\begin{align*}
q_1 & \begin{cases} \dot{x} = 0.0403x+0.5689y  \\
                             \dot{y} = 0.6771x-0.2556y
       \end{cases}\\
q_2 & \begin{cases} \dot{x} = 0.2617x-0.2747y  \\
                             \dot{y} = 1.2134x-0.1331y
       \end{cases}\\
q_3 & \begin{cases} \dot{x} = 1.4725x-1.2173y  \\
                             \dot{y} = 0.0557x-0.0412y
       \end{cases}\\
q_4 & \begin{cases} \dot{x} = -0.5217x+0.8701y  \\
                             \dot{y} = -1.4320x+0.8075y
       \end{cases}\\
q_5 & \begin{cases} \dot{x} = -2.1707x-1.0106y  \\
                             \dot{y} = -0.0592x+0.6145y
       \end{cases}
\end{align*}
The region $S$ is $[-1 \ \ 1]^2$, $\sigma_{I} = 0.5$ and $\sigma_G = 0.05$.
\end{system}

\begin{system}
\label{sys:dc-motor}This system is adopted from ~\cite{Mazo+Others/2010/PESSOA}
is a DC motor system. There are two continuous variables $\omega$ and
 $i$, and input $u$ is the source voltage.
\begin{align*}
    \dot{\omega} & = - \frac{B}{J}\omega + \frac{k}{J} i \\
    \dot{i} & = - \frac{k}{L} \omega - \frac{R}{L} i + \frac{1}{L}u
\end{align*}
, where $B = 10^{-4}$, $J = 25\times 10^{-5}$, $k = 0.05$, $R = 0.5$, 
$L = 15\times 10^{-4}$ and $u \in \{-1, 1\}$. The desired point is 
$[\omega \ i] = [20 \ 0 ]$ and by change of basis along with scaling,
the following system is obtained

\begin{align*}
    \dot{\omega'} & = - \frac{B}{J}(\omega' + 2) + \frac{k}{J} i \\
    \dot{i}' & = - \frac{k}{L} (\omega' + 2) - \frac{R}{L} i + 0.1 \frac{1}{L}u
\end{align*}
Region of interest $S = \{[\omega \ \ i]^T | \omega \in [-1 \ \ 1], i \in [-1 \ \ 1]\}$
, $\sigma_{I} = 0.4$ and $\sigma_G = 0.05$.
\end{system}

\begin{system}
\label{sys:inverted-pendulum}This system is adopted from ~\cite{PESSOA:Website}
is a model of inverted pendulum on a cart.
 There are two continuous variables $\theta$ (angular position)and
 $\omega$ (angular velocity), and input $u$ is the applied force to the cart.
\begin{align*}
\dot{\theta} & =\omega\\
\dot{\omega} & =\frac{g}{l}sin(\theta)-\frac{h}{ml^2}\omega+\frac{1}{ml}cos(\theta)u
\end{align*}
, where $g = 9.8$, $h = 2$, $l = 2$, $m = 0.5$ and $u \in \{-30, 30\}$. The region is 
$S = \{[\theta \ \ \omega]^T | \theta \in [-1.5 \ \ 1.5], i \in [-4 \ \ 4]\}$
, $\sigma_{I} = 0.5$ and $\sigma_G = 0.2$.
\end{system}

\begin{system}
\label{sys:dc-dc}
This system is a DCDC boost converter adopted from 
~\cite{mouelhi2013cosyma} with two discrete mode ($q_1$, $q_2$),
two continuous variables $i$ and $v$. By a simple change of bases the
state $i = 1.35$ and $v = 5.65$ is set as desired point of activity (origin)
and the following dynamics are obtained.
\begin{align*}
 q_1 & \begin{cases} \dot{i} = 0.0167i + 0.3333  \\ 
                             \dot{v} = - 0.0142v
       \end{cases}\\
 q_2 & \begin{cases} \dot{i} = - 0.0183i - 0.0663v + 0.3333  \\ 
                             \dot{v} = 0.0711i - 0.0142v
       \end{cases}\\
\end{align*}
Region of interest is $S = \{[i \ \ v]^T | i \in [-0.7 \ \ 0.7], v \in [-0.7 \ \ 
0.7]\}$, $\sigma_{I} = 0.3$ and $\sigma_G = 0.04$. Notice region $S$ is a little
different from the one explained in~\cite{mouelhi2013cosyma}.

\end{system}

\begin{system}
\label{sys:tulip-2d}
This system is adapted from~\cite{nilssonincremental}. There are two
continuous variables $x_1$ and $x_2$ and the controller can choose between
three different modes ($q_1$, $q_2$). By setting $x_1 = -0.75$ and $x_2 = 1.75$
as the origin, the new dynamics for these modes are
\begin{align*}
 q_1 & \begin{cases} \dot{x_1} = - x_2 -1.5 x_1 - 0.5 x_1^3  \\ 
                             \dot{x_2} = x_1 - x_2^2 + 2
       \end{cases}\\
 q_2 & \begin{cases} \dot{x_1} = - x_2 -1.5 x_1 - 0.5 x_1^3  \\ 
                             \dot{x_2} = x_1 - x_2
       \end{cases}\\
  q_3 & \begin{cases} \dot{x_1} = - x_2 -1.5 x_1 - 0.5 x_1^3 + 2 \\ 
                             \dot{x_2} = x_1 + 10
       \end{cases}
\end{align*}
Region $S$ is defined as $S = \{[x_1 \ \ x_2]^T | x_1 \in [-2.25 \ \ 2.75], v \in [-3.25 \ \ 3.25]\}$
, $\sigma_{I} = 1.0$ and $\sigma_G = 0.25$. 
Notice that this region is a little different from the one introduced in~\cite{nilssonincremental}.
Also, we scale the problem s.t.
$S = \{[x_1 \ \ x_2]^T | x_1 \in [-0.45 \ \ 0.45], v \in [-0.65 \ \ 0.65]\}$.
\end{system}

\begin{system}
\label{sys:heater-3d}
The system is a heater for keeping several rooms warm~\cite{mouelhi2013cosyma}.
There are 3 rooms $t_1$, $t_2$ and $t_3$ and heater can be in one of these
room or it can be off. Therefore, there are four modes ($q_0,...,q_3$) with the
following dynamics. The goal is to keep $t_i$ around 21 ($i \in \{1, 2, 3\}$).
	\begin{small}
\begin{align*}
q_0 & \begin{cases} \dot{t_1} = 0.01 (- 10.5(t_1+21) + 5(t_2+21) + 5(t_3+21) + 5)  \\ 
                    \dot{t_2} = 0.01 (5(t_1+21) - 10.5(t_2+21) + 5(t_3+21) + 5)  \\ 
                    \dot{t_3} = 0.01 (5(t_1+21) + 5(t_2+21) - 10.5(t_3+21) + 5)
       \end{cases}\\
q_1 & \begin{cases} \dot{t_1} = 0.01 (- 11.5(t_1+21) + 5(t_2+21) + 5(t_3+21) + 55)  \\ 
                    \dot{t_2} = 0.01 (5(t_1+21) - 10.5(t_2+21) + 5(t_3+21) + 5)  \\ 
                    \dot{t_3} = 0.01 (5(t_1+21) + 5(t_2+21) - 10.5(t_3+21) + 5)
       \end{cases}\\
q_2 & \begin{cases} \dot{t_1} = 0.01 (- 10.5(t_1+21) + 5(t_2+21) + 5(t_3+21) + 5)  \\ 
                    \dot{t_2} = 0.01 (5(t_1+21) - 11.5(t_2+21) + 5(t_3+21) + 55)  \\ 
                    \dot{t_3} = 0.01 (5(t_1+21) + 5(t_2+21) - 10.5(t_3+21) + 5)
       \end{cases}\\
q_3 & \begin{cases} \dot{t_1} = 0.01 (- 10.5(t_1+21) + 5(t_2+21) + 5(t_3+21) + 5)  \\ 
                    \dot{t_2} = 0.01 (5(t_1+21) - 10.5(t_2+21) + 5(t_3+21) + 5)  \\ 
                    \dot{t_3} = 0.01 (5(t_1+21) + 5(t_2+21) - 11.5(t_3+21) + 55)
       \end{cases}
\end{align*}
	\end{small}
Region $S = [-5 \ \ 5]^3$, $\sigma_{I} = 2.5$ and $\sigma_G = 1$.
\end{system}

\begin{system}
\label{sys:non-equilibrium-stabilization}
This system with $3$ continuous variables and 4 modes is adopted 
from~\cite{bolzern2004quadratic}. The dynamics are

\begin{align*}
q_1 & \begin{cases} \dot{x} = 4.15x - 1.06y - 6.7z + 1  \\ 
                             \dot{y} = 5.74x+4.78y-4.68z -4\\
                             \dot{z} = 26.38x-6.38y-8.29z+1
       \end{cases}\\
q_2 & \begin{cases} \dot{x} = -3.2x -7.6y -2z +4  \\ 
                             \dot{y} = 0.9x + 1.2y -z -2 \\
                             \dot{z} = x + 6y +5z -1
       \end{cases}\\
q_3 & \begin{cases} \dot{x} = 5.75x -16.48y -2.41z -2  \\ 
                             \dot{y} = 9.51x -9.49y +19.55z +1 \\
                             \dot{z} = 16.19x + 4.64y +14.05z -1
       \end{cases}\\
q_4 & \begin{cases} \dot{x} = -12.38x +18.42y +0.54z -1  \\ 
                             \dot{y} = -11.9x +3.24y -16.32z +2 \\
                             \dot{z} = -26.5x -8.64y -16.6z +1
       \end{cases}
\end{align*}
Region of interest is $P = [-1 \ \ 1]^3$, $\sigma_{I} = 0.5$ and $\sigma_G = 0.1$.

\end{system}

\begin{system}
\label{sys:linear-ss-2}
The system is a linear switched system, adapted 
from~\cite{pettersson2001stabilization}.
There are three continuous variables $x$, $y$, $z$ in this system 
and the dynamics for $3$ modes ($q_1$, $q_2$ and $q_3$) are
\begin{align*}
q_1 & \begin{cases} \dot{x} = 1.8631x - 0.0053y + 0.9129z  \\ 
                             \dot{y} = 0.2681x - 6.4962y + 0.0370z \\
                             \dot{z} = 2.2497x - 6.7180y + 1.6428z
       \end{cases}\\
q_2 & \begin{cases} \dot{x} = - 2.4311x - 5.1032y + 0.4565z  \\ 
                             \dot{y} = - 0.0869x + 0.0869y + 0.0185z \\
                             \dot{z} = 0.0369x - 5.9869y + 0.8214z
       \end{cases}\\
q_3 & \begin{cases} \dot{x} = 0.0372x - 0.0821y - 2.7388z  \\ 
                             \dot{y} = 0.1941x + 0.2904y - 0.1110z \\
                             \dot{z} =  - 1.0360x + 3.0486y - 4.9284z
       \end{cases}
\end{align*}
Region $S = [-1 \ \ 1]^3$, $\sigma_{I} = 0.3$ and $\sigma_G = 0.01$.
\end{system}

\begin{system}
\label{sys:linear-ss-3}
This system is a switched system
adopted from ~\cite{greco2005stability}.
There are three continuous variables $x$, $y$, $z$ and
$5$ modes ($q_1,..., q_5$) the dynamics of each mode 
is described below
\begin{align*}
q_1 & \begin{cases} \dot{x} = 0.1764x + 0.8192y - 0.3179z  \\ 
                             \dot{y} = -1.8379x-0.2346y-0.7963z \\
                             \dot{z} = -1.5023x-1.6316y+0.6908z
       \end{cases}\\
q_2 & \begin{cases} \dot{x} = -0.0420x-1.0286y+0.6892z  \\ 
                             \dot{y} = 0.3240x+0.0994y+1.8833z \\
                             \dot{z} = 0.5065x-0.1164y+0.3254z
       \end{cases}\\
q_3 & \begin{cases} \dot{x} = -0.0952x-1.7313y+0.3868z  \\ 
                             \dot{y} = 0.0312x+0.4788y+0.0540z \\
                             \dot{z} = -0.6138x-0.4478y-0.4861z
       \end{cases}\\
q_4 & \begin{cases} \dot{x} = 0.2445x+0.1338y+1.1991z  \\ 
                             \dot{y} = 0.7183x-1.0062y-2.5773z \\
                             \dot{z} = 0.1535x+1.3065y-2.0863z
       \end{cases}\\
q_5 & \begin{cases} \dot{x} = -1.4132x-1.4928y-0.3459z  \\ 
                             \dot{y} = -0.5918x-0.0867y+0.9863z \\
                             \dot{z} = 0.5189x-0.0126y+0.6433z
       \end{cases}
\end{align*}
Region $S = [-3 \ \ 3]^3$, $\sigma_{I} = 1$ and $\sigma_G = 0.1$.
\end{system}

\begin{system}
\label{sys:lorenz}
This system is adopted from ~\cite{saat2011nonlinear}.
There are three continuous variables $x$, $y$, $z$ and the dynamics are
\begin{align*}
	\dot{x} &= -10x + 10y + u \\
	\dot{y} &=  28x - y -xz \\
	\dot{z} &=  xy - 2.6667z
\end{align*}
, where $u \in \{-100, 100\}$. And region $S = [-5 \ \ 5]^3$, 
$\sigma_{I} = 1.2$ and $\sigma_G = 0.3$.
\end{system}

\begin{system}
\label{sys:tulip-pipe-3d}
This system is a radiant system in building adopted from 
~\cite{nilssonincremental} which is a switched linear system with three
continuous variables ($T_c$, $T_1$ and $T_2$) and two modes
 ($q_1$, $q_2$). By setting $T_c = 24$ and $T_1 = T_2 = 23$
 as the new origin, the dynamics obtained are

 \begin{align*}
 q_1 & \begin{cases} \dot{T_c} = 2.25T_1 + 2.25T_2 - 9.26T_c - 14.54  \\ 
                             \dot{T_1} = 2.85T_2 -7.13T_1 + 4.04T_c + 4.04 \\
                             \dot{T_2} = 2.85T_1- 7.13T_2 + 4.04T_c + 4.04
       \end{cases}\\
 q_2 & \begin{cases} \dot{T_c} = 2.25T_1 + 2.25T_2 - 4.5T_c + 4.5  \\ 
                             \dot{T_1} = 2.85T_2 -7.13T_1 + 4.04T_c + 4.04 \\
                             \dot{T_2} = 2.85T_1- 7.13T_2 + 4.04T_c + 4.04
       \end{cases}
\end{align*}
Region $S = [-6 \ \ 6]^3$, $\sigma_{I} = 3$ and $\sigma_G = 1$.
\end{system}

\begin{system}
\label{sys:heater-4d}
The system is similar to System~\ref{sys:heater-3d}, except that the number of dimensions is 4.
 See~\cite{mouelhi2013cosyma}.
\end{system}

\begin{system}
\label{sys:LQR}
The original system is a switched control system with inputs 
from~\cite{zhang2009exponential}. There are $4$ variables ($w$, $x$ ,$y$ and $z$) 
and $4$ original modes. After converting the discrete system into a continuous one, the dynamics
are 

\begin{small}
\begin{align*}
 q_1 & \begin{cases} \dot{w} &= -0.693w   -1.099x    +2.197y    +3.296z -7.820u  \\ 
                     \dot{x} &= -1.792x    +2.197y    +4.394z   -8.735u \\
                     \dot{y} &= -1.097x    +1.504y    +2.197z   -2.746u\\
                     \dot{z} &= 0.406z    +3.244u
       \end{cases}\\
 q_2 & \begin{cases} \dot{w} &= -1.792w   -1.099x    +2.197y    +1.099z    +6.696u  \\ 
                     \dot{x} &= 0.406x   -2.197y     +4.734u \\
                     \dot{y} &= -0.693y    +2.773u\\
                     \dot{z} &= -2.197w   -1.099x    +2.197y    +1.504z    
                     +4.263u
       \end{cases}\\
  q_3 & \begin{cases} \dot{w} &= 0.406w    +0.811u  \\ 
                     \dot{x} &= 1.099w   -0.144x    +0.549y   -0.549z    +1.910u \\
                     \dot{y} &= 0.549x   -0.144y   -0.549z    +3.871u\\
                     \dot{z} &= 1.099w   -0.693z    +4.970u
       \end{cases}\\
  q_4 & \begin{cases} \dot{w} &= -0.693w    +2.000x    +1.863u  \\ 
                     \dot{x} &= -0.693x    +4.159u \\
                     \dot{y} &= -0.693y    +2.773u\\
                     \dot{z} &= 4.000x   -4.000y   -0.693z   -1.069u
       \end{cases}
\end{align*}
\end{small}
, where $u \in \{-1, 1\}$ and Region of interest is $S = [-1, 1]^4$, 
$\sigma_{I} = 0.1$ and $\sigma_G = 0.1$.
\end{system}

\begin{system}
\label{sys:Tora}
This system is a Tora system and the equations are adopted
from~\cite{faubourg1999design}. There are 4 variables
in this system with the following dynamics
\begin{align*}
	\dot{w} &= x \\
	\dot{x} &= -w + 0.1 \sin(y) \\
	\dot{y} &=  z \\
	\dot{z} &=  u
\end{align*}
, where $u \in [-10, 10]$ and region $S = [-1, 1]^4$, $\sigma_{I} = 0.1$ and 
$\sigma_G = 0.02$.
\end{system}

\begin{system}
\label{sys:heater-5d}
The system is similar to System~\ref{sys:heater-3d}, except that the number of dimensions is 5.
 See~\cite{mouelhi2013cosyma}.
\end{system}

\begin{system}
	\label{sys:heater-6d}
	This system is $6$ variables version of System~\ref{sys:heater-3d} and there are
	$6$ rooms and $2$ heaters and only consider $4$ modes considered. The heater is off for 
	one mode ($q_0$) and for mode $q_i$ ($1 \leq i \leq 3$), two heaters are on in
	rooms $i$ and $3+i$.
\end{system}

\begin{system}
	\label{sys:heater-9d}
	This system is $9$ variables version of System~\ref{sys:heater-3d} and there are
	$9$ rooms and $3$ heaters and only consider $4$ modes considered. The heater is off for 
	one mode ($q_0$) and for mode $q_i$ ($1 \leq i \leq 3$), three heaters are on in
	rooms $i$, $3+i$ and $6+i$.
\end{system}

\end{document}